\documentclass[prd, preprint, nofootinbib]{revtex4}
\usepackage{graphicx}
\usepackage{amsfonts}
\usepackage{latexsym}
\usepackage{amssymb}
\usepackage{epsfig}

\begin{document}

\title{ Polynomial inflation models after BICEP2 
}

\author{Tatsuo Kobayashi}
 \affiliation{Department of Physics, Hokkaido University, Sapporo 060-0810, Japan
}

\author{Osamu Seto}
 \affiliation{Department of Life Science and Technology,
  Hokkai-Gakuen University, Sapporo 062-8605, Japan
}

%
\begin{abstract}
Large field inflation models are favored by the recent BICEP2 
 that has detected gravitational wave modes generated during inflation.
We study general large field inflation models for which
 the potential contains
 (constant) quadratic and quartic terms of inflaton field.
We show, in this framework,
 those inflation models can generate the fluctuation with
 the tensor-to-scalar ratio of $0.2$ as well as the scalar spectral index of $0.96$:
 those are very close to the center value of the tensor-to-scalar ratio 
 reported by BICEP2 as well as Planck. 
Finally, we briefly discuss the particle physics model building of inflation.
\end{abstract}

\preprint{EPHOU-14-005}
\preprint{HGU-CAP-031}

\vspace*{3cm}
\maketitle


\section{Introduction}

The inflationary cosmological model is the standard paradigm of modern cosmology
 because an inflationary expansion in the very early Universe
 solves various problems in the standard big bang cosmology~\cite{Inflation}
 and also 
 provides the seed of large scale structure in our Universe from
 the quantum fluctuation of an inflaton field $\phi$~\cite{InflationFluctuation}.
The property of the generated density fluctuation from
 a single-field slow-roll inflation model,
 namely adiabatic, Gaussian, and its almost scale-invariant spectrum,
 is quite consistent with various cosmological observations.

As the scalar perturbation is generated from the inflaton's quantum fluctuation
 during inflationary expansion,
 the tensor perturbation also is generated from graviton's one~\cite{Tensor}.
The tensor perturbation induces $B$-mode polarization of 
 the temperature anisotropy in the cosmic microwave background radiation
 and is important for inflationary cosmology
 because the tensor perturbation directly tells us the energy scale of inflation.

Recently, the BICEP2 collaboration reported the detection of the tensor mode
 through the $B$-mode polarization with
 the corresponding tensor-to-scalar ratio~\cite{Ade:2014xna}
\begin{equation}
 r_T = 0.20^{+0.07}_{-0.05} .
\label{rT:BICEP}
\end{equation}
Its cosmological implications also have been studied~\cite{AfterBICEP}. 
It has been well known that theoretically such a large tensor-to-scalar ratio can
 be generated only by so-called large field models, where
 its potential, as of a polynomial function of $\phi$, is convex and
 the variation of the inflaton field value during inflation $\Delta\phi$ is 
 as large as of the Planck scale~\cite{LythBound}.
On the other hand, the Planck satellite~\cite{Ade:2013zuv,Ade:2013uln}
 has reported the scalar spectral index as
\begin{equation}
 n_s \simeq 0.96 .
\label{ns:Planck}
\end{equation}

Now, if we compare the central values of Eqs.~(\ref{rT:BICEP}) and (\ref{ns:Planck}) 
 with the predicted values of well-studied potential models
 such as $V \propto \phi^2 $ or $\phi^4$,
 there is a discrepancy.
Namely, $r_T$ from $V \propto \phi^2 $ is too low and that from $\phi^4$ is
 too high~\cite{Ade:2013uln}.
~\footnote{The $V \propto \phi^3$ potential would well agree with data. 
 However, naively, this potential is pathological because the potential is not
 bounded from below. Note, however, an effective realization would be possible 
 with a field redefinition from ${\cal L} \sim \phi^2(\partial\phi)^2 - \phi^6$. }

In this paper, we extend analysis to general polynomial models of inflation 
 for which the potential is expressed as~\cite{Poly}
\begin{equation}
 V = c_1 + c_2 \phi^2 +c_4 \phi^4 , 
\label{potential:general}
\end{equation}
 to examine whether this form of potential can reconcile the mismatch mentioned above, 
 by taking the latest BICEP2 data into account.
Since it is not easy to control so many parameters, 
 in practice, we will consider terms up to $\phi^4$,
 which might be motivated by the renormalizability of quantum field theory.

\section{Polynomial potential model}

We study canonical single field inflation models with a polynomial potential.
Within this framework, the power spectrum of the density perturbation,
 its spectral index and the tensor-to-scalar ratio are expressed as
\begin{eqnarray}
{\cal P}_{\zeta} &=& \left(\frac{H^2}{2\pi |\dot{\phi}|}\right)^2
 = \frac{V}{24 \pi^2 \epsilon}, \\
n_s &=& 1 + 2 \eta -6 \epsilon ,\\
r_T &=& 16 \epsilon ,
\end{eqnarray}
 respectively, by using slow-roll parameters
\begin{eqnarray}
 \eta &=& \frac{V_{\phi\phi}}{V} , \\
 \epsilon &=& \frac{1}{2}\left(\frac{V_{\phi}}{V}\right)^2 , 
\end{eqnarray}
 in the unit with $8 \pi G =1$.
Here,
 a subscript $\phi$ and dot denote a derivative with respect to $\phi$
 and time, respectively,
 and $H$ is the Hubble parameter of the Universe.

\subsubsection{Positive quadratic and quatic}

The first example is the inflation driven by the potential
\begin{equation}
 V = \frac{1}{2}m^2 \phi^2 + \frac{\lambda}{4} \phi^4 .
\end{equation}
For this potential, the slow-roll parameters are given by
\begin{eqnarray}
 \eta &=& 
 \frac{  m^2 + 3 \lambda \phi^2}{  \frac{1}{2}m^2 \phi^2 + \frac{\lambda}{4} \phi^4} , \\
 \epsilon &=& \
 \frac{1}{2}
 \left(\frac{  m^2 \phi + \lambda \phi^3}{  \frac{1}{2}m^2 \phi^2 + \frac{\lambda}{4} \phi^4}\right)^2 .
\end{eqnarray}
When the inflaton reaches $\phi^2=2$, then $\eta$ becomes unity, and inflation ends.
By solving the slow-roll equation $3 H \dot{\phi}+ V_{\phi}=0$, we obtain
\begin{eqnarray}
 N = \int_{\phi_e}^{\phi} \frac{V}{V_{\phi}} d\phi 
   \simeq \frac{\phi^2}{8}
 + \frac{m^2}{8 \lambda}\ln\left(\frac{m^2+\lambda \phi^2}{m^2}\right), 
\label{Sol:phi1}
\end{eqnarray}
 with $N$ being the number of $e$-folds during inflation,
 where $\phi_e=\sqrt{2}$ is the field value when inflation ends
 and is dropped in Eq.~(\ref{Sol:phi1}) due to the smallness compared to the others. 

\begin{figure}[!t]
\begin{center}
\epsfig{file=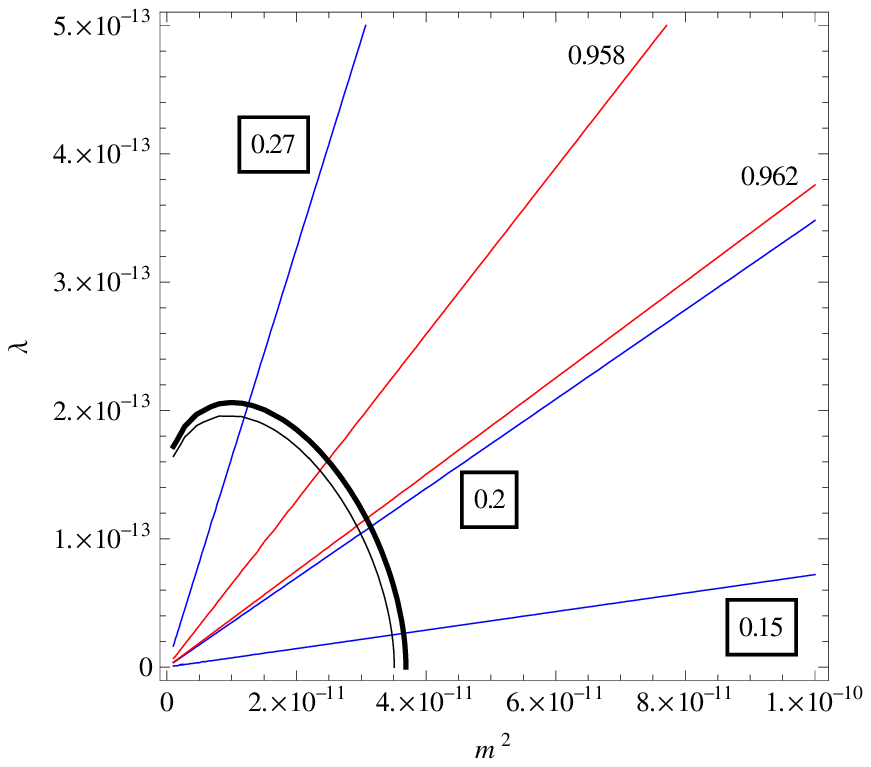, width=8cm,height=8cm,angle=0}
\end{center}
\caption{ Various contours for $N=60$; the
black double curve is for the amplitude of the density perturbation, 
blue lines are for the tensor-to-scalar ratio,
and red lines are the spectral index $n_s$. 
To reproduce the central value $r_T \simeq 0.2$,
 we need $\lambda \simeq 1 \times 10^{-13}$.
 }
\label{Fig:positive60}
%
\begin{center}
\epsfig{file=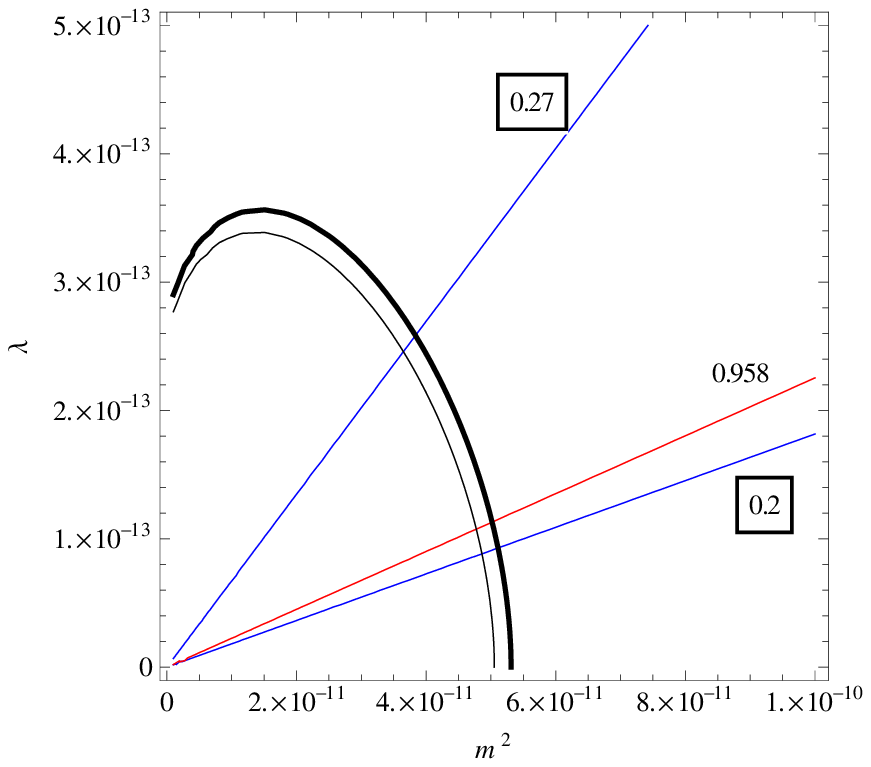, width=8cm,height=8cm,angle=0}
\end{center}
\caption{
Same as Fig.~\ref{Fig:positive60} but for $N=50$. 
To reproduce the central value $r_T \simeq 0.2$, we need $\lambda \simeq 0.8 \times 10^{-13}$.
 }
\label{Fig:positive50}
\end{figure}

We show the contours of inflationary observables in the $m^2 - \lambda $ plane. 
Figure~\ref{Fig:positive60} is for the case that
 the observed cosmological scale $k_*=0.002$ Mpc${}^{-1}$ is assumed to
 correspond to $N=60$.
The double curve depicts the amplitude of the power spectrum of the density perturbation
 ${\cal P}_{\zeta} = (22.42-21.33)\times 10^{-10}$ quoted from Table 3 ($\Lambda$CDM)
 in Ref.~\cite{Ade:2013uln}.
Red lines with numbers are contours of $n_s$.
Blue lines with numbers enclosed by a square are contours of $r_T$,
 for which the values are those noted in Eq.~(\ref{rT:BICEP}).  
For $N=60$, the predicted $r_T$ without $\lambda$ lies outside
 of the error bar reported by BICEP2.
However, including a small quartic term with $\lambda \simeq 1\times 10^{-13}$,
 we obtain $r_T \simeq 0.2$ and $n_s \simeq 0.96$.
Figure~\ref{Fig:positive50} is for $N=50$ and shows that
 $\lambda \simeq 0.8\times 10^{-13}$ provides the best fitting to $r_T$.

\subsubsection{Negative quadratic and quartic}

Next, let us consider the case with a negative quadratic term;
 in other words, this is a double-well potential.
We introduce a constant term as well in order to
 realize the vanishing cosmological constant at the true minimum. 
The potential is given by
\begin{equation}
 V = V_0 - \frac{1}{2}m^2 \phi^2 + \frac{\lambda}{4} \phi^4 .
\end{equation}
From the stationary condition and the vanishing cosmological constant condition 
 at the minimum, we find the vacuum expectation value (VEV) of $\phi$ and $V_0$ as
\begin{eqnarray}
&& \langle\phi\rangle^2 = \frac{m^2}{\lambda}, \\
&& V_0 =  \frac{m^4}{4 \lambda}.
\end{eqnarray}
The slow-roll parameters are given by
\begin{eqnarray}
 \eta &=& 
 \frac{ -m^2 + 3 \lambda \phi^2}{V_0-\frac{1}{2}m^2 \phi^2 + \frac{\lambda}{4} \phi^4} , \\
 \epsilon &=& \
 \frac{1}{2}
 \left(\frac{ -m^2 \phi + \lambda \phi^3}{V_0-\frac{1}{2}m^2 \phi^2 + \frac{\lambda}{4} \phi^4}\right)^2 . 
\end{eqnarray}
$\eta$ becomes unity and inflation ends when the inflaton reaches
\begin{equation}
\phi_e^2 = 6 + \frac{ m^2 +2 \sqrt{\lambda(9\lambda + 2 m^2)} }{\lambda}.
\end{equation}
By solving the slow-roll equation, we obtain
\begin{eqnarray}
 N = \int_{\phi_e}^{\phi} \frac{V}{V'} d\phi 
  \simeq  \frac{\phi^2-\phi_e^2}{8}
 - \frac{m^2}{8 \lambda}\ln\left(\frac{\phi^2}{\phi_e^2}\right) .
\label{Sol:phi2}
\end{eqnarray}

\begin{figure}[!t]
\begin{center}
\epsfig{file=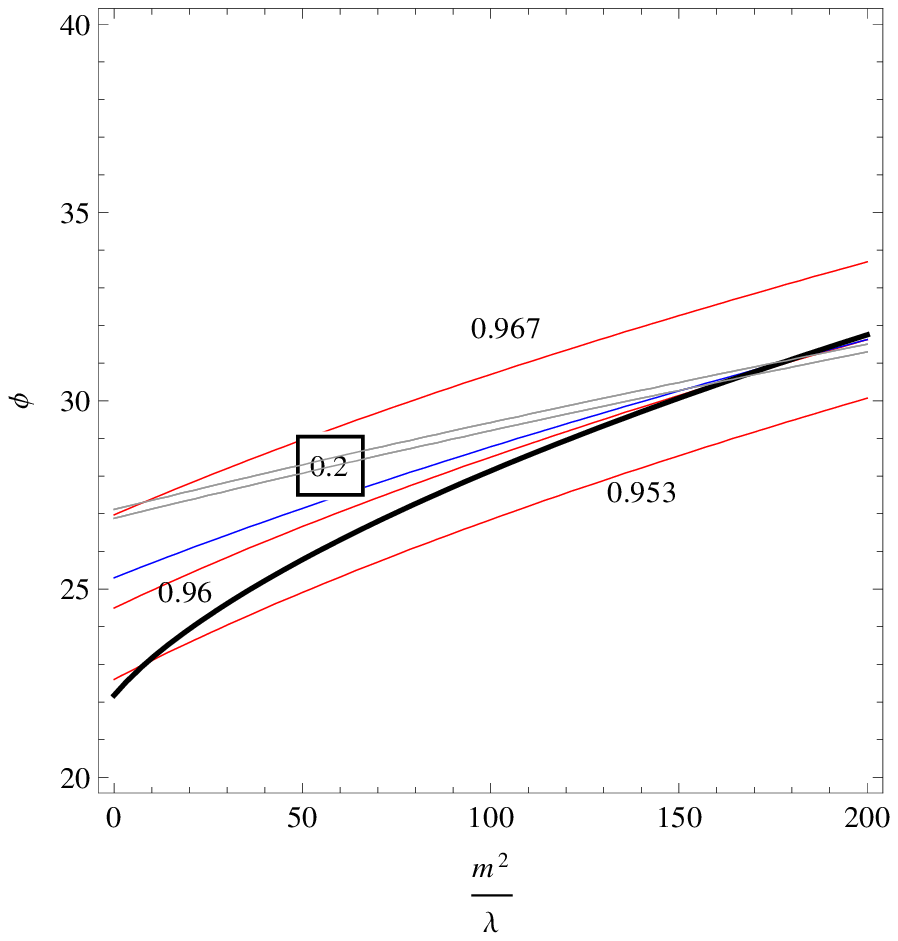, width=8cm,height=8cm,angle=0}
\end{center}
\caption{ Various contours for $N=60$;
 the black double curve is for the amplitude of the density perturbation with $\lambda = 4.5\times 10^{-14}$, 
the blue line is for the tensor-to-scalar ratio,
and the red lines are the spectral index $n_s$. 
To reproduce the central value $r_T \simeq 0.2$, we need $m^2/\lambda =\langle\phi\rangle^2  \simeq 170$.
 }
\label{Fig:negative60}
%
\begin{center}
\epsfig{file=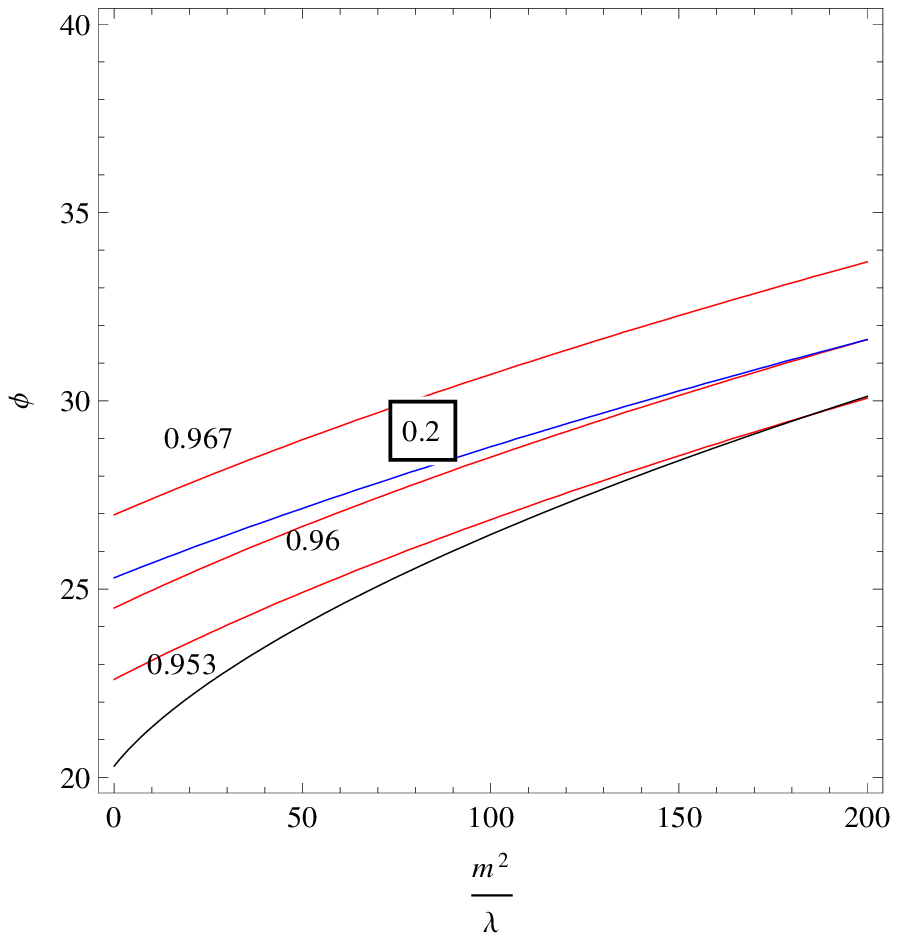, width=8cm,height=8cm,angle=0}
\end{center}
\caption{
Same as Fig.~\ref{Fig:negative60} but for $N=50$. 
To reproduce the central value $r_T \simeq 0.2$, we need $m^2/\lambda =\langle\phi\rangle^2  \simeq {\cal O}(10^3)$. 
 }
\label{Fig:negative50}
\end{figure}

We show the contours of inflationary observables in the $m^2/\lambda - \phi $ plane,
 with $\phi$ being the field value during inflation given in Eq.~(\ref{Sol:phi2}). 
Red lines with numbers are contours of $n_s$ with the error band reported by Planck.
The blue line with the number $0.2$ enclosed by
 a square is the central value contour of $r_T$ by BICEP2.  
Without the $m$ term, in other words in the $V=\phi^4$ potential limit,
 the predicted $n_s$ lies outside of the uncertainty band reported by Planck,
 as is well known~\cite{Ade:2013uln}.
The black thick and solid line correspond to the $\phi$ field value for $N=60$ and $50$
 in Figs.~\ref{Fig:negative60} and \ref{Fig:negative50}, respectively.
As increasing $m$, those lines approach observed values of $n_s$ as well as $r_T$.
From Fig.~\ref{Fig:negative60}, we find
 that the $r_T=0.2$ contour meets the $\phi(N=60)$ line
 at $m^2/\lambda \simeq 170$, which corresponds to
\begin{equation}
 \langle\phi\rangle \simeq 13.
\end{equation}
The double-well potential with this VEV well reproduces those observed values.
The double curve for ${\cal P}_{\zeta} = (22.42-21.33)\times 10^{-10}$
 is drawn with $\lambda = 4.5\times 10^{-14}$ in Fig.~\ref{Fig:negative60}.

Figure~\ref{Fig:negative50} is for $N=50$,
 where we omit the ${\cal P}_{\zeta}$ contour for simplicity.
We need $m^2/\lambda = {\cal O}(10^3)$ to realize $r_T=0.2$
 for $N=50$, which is far outside of the range in Fig.~\ref{Fig:negative50}.

\section{Summary and discussion}

We have shown that general polynomial inflation models are fit well to
 the observed data including the tensor-to-scalar ratio recently reported by BICEP2.
In other words, after we know the size of $r_T$, 
 we are able to determine more parameters of inflation models.
In fact, for $V \sim + \phi^2 + \phi^4$ with $N=60$,
 we find $m^2 \simeq 3\times 10^{-11}$ and $\lambda \simeq 1\times 10^{-14}$.
A double-well potential also can be fit nicely to the data; then,
 the VEV should be $\langle\phi\rangle \simeq 13$, and
 the self-coupling constant is $\lambda \simeq 4.5\times 10^{-14}$ for $N=60$.

Finally, we note here possible directions of particle physics model construction
 for inflation.
Being aware of the above size of self-coupling,
 an inflaton's Yukawa coupling $y$ to fermion $\psi$, ${\cal L} = y \bar{\psi}\phi\psi$,
 should be smaller than ${\cal O}(10^{-4})$
 because it induces a ${\cal O}(y^4)$ self-coupling by radiative corrections.
Supersymmetric construction of a large field inflation model has been
 challenging~\cite{Yamaguchi:2011kg}.
F-term inflation suffers from so-called ``$\eta$ problem''
 due to higher-order terms from the Kahler potential.
Imposing shift symmetry $\phi \rightarrow \phi + C$
 is one of a few effective manners to overcome the problem~\cite{Kawasaki:2000yn,Kaloper:2008fb}.
While D-term inflation has been regarded as an example of hybrid inflation~\cite{Binetruy:1996xj},
 in fact, D-term chaotic inflation is also possible~\cite{Kadota:2007nc}.
Such a model interestingly does not suffer from the $\eta$ problem
 even if we consider a general Kahler potential
 because the field redefinition to the canonical field absorbs
 higher-order corrections and
 the Lagrangian is reduced to a quartic or double-well potential.
Since a self-coupling is given by the square of a gauge coupling in the D-term,
 we need to introduce a new gauge interaction with the
 gauge coupling constant of ${\cal O}(10^{-7})$.
If we could realize a large VEV about $10$ by any means, as we have shown,
 such a model would easily reproduce the data.

We have restricted the polynomial potential 
(\ref{potential:general}) in this paper.
It would be important to add higher-order terms $\phi^n$ 
as well as the cubic term $\phi^3$.
We will study it elsewhere.


\section*{Acknowledgments}
T.K. was supported in part by the Grant-in-Aid for Scientific Research 
No.~25400252 from the Ministry of Education, Culture, Sports, Science 
and Technology in Japan. 
%




\begin{thebibliography}{99}


\bibitem{Inflation}
  A.~A.~Starobinsky,
  JETP Lett.\  {\bf 30} 682 (1979)
  [Pisma Zh.\ Eksp.\ Teor.\ Fiz.\  {\bf 30} 719 (1979)];\\
  K.~Sato,
  Mon.\ Not.\ R.\ Astron.\ Soc.\  {\bf 195}, 467 (1981);\\
  A.~H.~Guth,
  Phys.\ Rev.\  D {\bf 23}, 347 (1981); \\
  A.~D.~Linde,
  Phys.\ Lett.\  B {\bf 108} 389 (1982);\\
  A.~Albrecht and P.~J.~Steinhardt,
  Phys.\ Rev.\ Lett.\  {\bf 48} 1220 (1982).

\bibitem{InflationFluctuation}
  S.~W.~Hawking,
  Phys.\ Lett.\  B {\bf 115}, 295 (1982); \\
  A.~A.~Starobinsky,
  Phys.\ Lett.\  B {\bf 117}, 175 (1982); \\
  A.~H.~Guth and S.~Y.~Pi,
  Phys.\ Rev.\ Lett.\  {\bf 49}, 1110 (1982).

\bibitem{Tensor} 
  A.~A.~Starobinsky,
  JETP Lett.\  {\bf 30}, 682 (1979)
  [Pisma Zh.\ Eksp.\ Teor.\ Fiz.\  {\bf 30}, 719 (1979)]; \\
  V.~A.~Rubakov, M.~V.~Sazhin and A.~V.~Veryaskin,
  Phys.\ Lett.\ B {\bf 115}, 189 (1982);\\
  L.~F.~Abbott and M.~B.~Wise,
  Nucl.\ Phys.\ B {\bf 244}, 541 (1984);\\
  B.~Allen,
  Phys.\ Rev.\ D {\bf 37}, 2078 (1988).

\bibitem{Ade:2014xna} 
  P.~A.~R.~Ade {\it et al.}  [BICEP2 Collaboration],
  arXiv:1403.3985 [astro-ph.CO].

\bibitem{AfterBICEP}
  J.~Joergensen, F.~Sannino and O.~Svendsen,
  arXiv:1403.3289 [hep-ph];\\
  K.~Nakayama and F.~Takahashi,
  arXiv:1403.4132 [hep-ph];\\
  K.~Harigaya, M.~Ibe, K.~Schmitz and T.~T.~Yanagida,
  arXiv:1403.4536 [hep-ph];\\
  L.~A.~Anchordoqui, V.~Barger, H.~Goldberg, X.~Huang and D.~Marfatia,
  arXiv:1403.4578 [hep-ph];\\
  M.~Czerny, T.~Kobayashi and F.~Takahashi,
  arXiv:1403.4589 [astro-ph.CO];\\
  Y.~-Z.~Ma and Y.~Wang,
  arXiv:1403.4585 [astro-ph.CO];\\
  H.~Collins, R.~Holman and T.~Vardanyan,
  arXiv:1403.4592 [hep-th];\\
  C.~R.~Contaldi, M.~Peloso and L.~Sorbo,
  arXiv:1403.4596 [astro-ph.CO];\\
  K.~Harigaya and T.~T.~Yanagida,
  arXiv:1403.4729 [hep-ph];\\
  A.~Kehagias and A.~Riotto, 
  arXiv:1403.4811 [astro-ph.CO];\\
  J.~Lizarraga, J.~Urrestilla, D.~Daverio, M.~Hindmarsh, M.~Kunz, A.~R.~Liddle,
  arXiv:1403.4924 [astro-ph.CO]. 
 
\bibitem{LythBound}
 D.~H.~Lyth, Phys.\ Rev.\ Lett.\ {\bf 78}, 1861 (1997).

\bibitem{Ade:2013zuv} 
  P.~A.~R.~Ade {\it et al.}  [Planck Collaboration],
  arXiv:1303.5076 [astro-ph.CO].
  
\bibitem{Ade:2013uln} 
  P.~A.~R.~Ade {\it et al.}  [Planck Collaboration],
  arXiv:1303.5082 [astro-ph.CO].

\bibitem{Poly} 
  For recent previous studies, see e.g.,
  J.~Martin, C.~Ringeval and V.~Vennin,
  arXiv:1303.3787 [astro-ph.CO];\\
  J.~Martin, C.~Ringeval, R.~Trotta and V.~Vennin,
  arXiv:1312.3529 [astro-ph.CO];\\
  K.~Nakayama, F.~Takahashi and T.~T.~Yanagida,
  Phys.\ Lett.\ B {\bf 725}, 111 (2013);
  JCAP {\bf 1308}, 038 (2013).

\bibitem{Yamaguchi:2011kg} 
  D.~H.~Lyth and A.~Riotto,
  Phys.\ Rept.\  {\bf 314}, 1 (1999);\\
  M.~Yamaguchi,
  Class.\ Quant.\ Grav.\  {\bf 28}, 103001 (2011).

\bibitem{Kawasaki:2000yn} 
  M.~Kawasaki, M.~Yamaguchi and T.~Yanagida,
  Phys.\ Rev.\ Lett.\  {\bf 85}, 3572 (2000).

\bibitem{Kaloper:2008fb} 
  N.~Kaloper and L.~Sorbo,
  Phys.\ Rev.\ Lett.\  {\bf 102}, 121301 (2009);\\
  N.~Kaloper, A.~Lawrence and L.~Sorbo,
  JCAP {\bf 1103}, 023 (2011).
  
\bibitem{Binetruy:1996xj} 
  P.~Binetruy and G.~R.~Dvali,
  Phys.\ Lett.\ B {\bf 388}, 241 (1996);\\
  E.~Halyo,
  Phys.\ Lett.\ B {\bf 387}, 43 (1996).

\bibitem{Kadota:2007nc} 
  K.~Kadota and M.~Yamaguchi,
  Phys.\ Rev.\ D {\bf 76}, 103522 (2007).


\end{thebibliography}
\end{document}